\documentclass[english,twocolumn,twocolumn,showpacs,preprintnumbers,amsmath,amssymb,superscriptaddress]{revtex4}
\usepackage[T1]{fontenc}
\usepackage[latin1]{inputenc}
\usepackage{graphicx}

\makeatletter

\makeatletter

\makeatletter

\makeatletter

\makeatletter

\usepackage{color}

\usepackage{dcolumn}

\usepackage{bm}

\newcommand{\pt}{ p_{\rm t}}
\newcommand{\ie}{{\it i.e.}}

 \def\lsim{\mathrel{\rlap{\lower4pt\hbox{\hskip1pt$\sim$}}
    \raise1pt\hbox{$<$}}}         
 \def\gsim{\mathrel{\rlap{\lower4pt\hbox{\hskip1pt$\sim$}}
    \raise1pt\hbox{$>$}}}         

\makeatother

\makeatother

\usepackage{babel}
\makeatother
\begin{document}
{\small \it SQM 2008 contribution}
\vspace{0.5in}
\title{Jet quenching and Direct Photon Production }

\author{Fu-Ming Liu }

\affiliation{Institute of Particle Physics, Central China Normal University, Wuhan,
China }

\author{Tetsufumi Hirano}

\affiliation{Department of Physics, The University of Tokyo, 113-0033, Japan}

\author{Klaus Werner}

\affiliation{Laboratoire SUBATECH, University of Nantes - IN2P3/CNRS - Ecole desMines,
Nantes, France }

\author{Yan Zhu}

\affiliation{Institute of Particle Physics, Central China Normal University, Wuhan,
China }

\date{\today}

\begin{abstract}
Jet quenching effect has been investigated in the direct photon production, based on a realistic data-constrained (3+1) dimensional hydrodynamic
description of the expanding hot and dense matter, a reasonable treatment
of the propagation of partons and their energy loss in the fluid,
and a systematic study of the main sources of direct photons. Our
resultant $\pt$ spectra agree with recent PHENIX data in a broad
$\pt$ range. 
Parton energy loss in the plasma eventually effect significantly 
direct photon production from fragmentation and jet photon conversion,
similar to hadron suppression in central heavy ion collisions.
But this only causes about 40\% 
decrease in the total production of direct photons, due to the mixture with other direct photon sources. 
\end{abstract}
\maketitle

\section{Introduction}

The formation and observation of a quark-gluon plasma (QGP) in heavy
ion collisions are important goals of modern nuclear physics~\cite{QM06,harris96}.
Suppression of high $\pt$ hadron yields~\cite{supression} is one
of the most important features observed at the Relativistic Heavy
Ion Collider (RHIC). Theoretically this is attributed to the interaction
between jets (hard partons) and the bulk matter \cite{bjorken82,Gy90,Baier97,BDMPS}.
This offers us an excellent opportunity to study the interaction of
partons inside the system and, consequently, properties of the matter
under extreme conditions. 

However the observed centrality-dependence of hadronic jet quenching
doesn't show up in direct photon production\cite{Adler:2006hu}. In this paper, we investigate
the role of jet quenching on direct photon production. For this purpose,
we need a careful consideration of three components: 1) the space-time
evolution of the hot dense matter or the bulk matter; 2) hard parton
production and propagation in the bulk and 3) all sources of direct
photons. 

The space-time evolution of the hot dense matter is achieved by using
three dimensional (3D) hydrodynamic simulations of bulk matter \cite{Hirano:2001eu,Hirano02}
which have already been tested against a vast body of low $\pt$ hadron
data at RHIC.

The interaction between hard parton and the bulk so that energy loss
of the hard parton is formulated via the BDMPS framework \cite{BDMPS}
and tested on pion suppression at various centralities. 

Four significant sources are considered in this paper: leading order
contribution in Primordial NN scattering, thermal contribution, contribution
from jet-photon conversion and from jet fragmentation. There are possible
sources ignored at present study: The medium-induced photon radiation
is expected to contribute at low transverse momentum region where
thermal contribution dominants. The interaction involving non-equilibrated
soft partons may also produce direct photons, but the preequilibrium
time interval is much shorter than the life time of the equilibrated
matter ($\sim20$ fm/$c$). 

Therefore energy loss of hard parton in the bulk will directly effect
two direct photon sources, jet-photon conversion and from jet fragmentation.
We will see the role of jet quenching on direct photon production
clearly via investigating those production in Au-Au collisions at
different centralities where various sizes of hot dense matter are
formed. Certainly different sizes of bulks will also contribute thermal
photon production differently which will not be discussed in detail
in this paper.

The paper is organized as follows: In Sec.~\ref{sec:2}, we first
give a brief review on the space-time evolution of the hot matter
created in Au+Au collisions at different centralities based on a (3+1)-dimensional
ideal hydrodynamical calculation. In Sec.~\ref{sec:3}, we discuss
parton energy loss in the QGP. We investigate neutral pion production
in the high $\pt$ region in order to fix the parameters of the energy
loss scheme. We discuss sequently the contributions from various sources
to direct photon $\pt$ spectra in Sec.~\ref{sec:4}. We show our
results and compare them with recent experimental data in Sec.~\ref{sec:5}.
Section~\ref{sec:6} is devoted to conclusion of the present study.

\section{space-time evolution of the hot and dense matter }

\label{sec:2}

In our calculation, fully three-dimensional (3D) ideal hydrodynamics
\cite{Hirano:2001eu,Hirano02} is employed to describe the space-time
evolution of the hot and dense matter created in Au+Au collisions
at RHIC energy at various centralities. We solve the equations of
energy-momentum conservation \begin{eqnarray}
\partial_{\mu}T^{\mu\nu}=0\end{eqnarray}
 in full 3D space $(\tau,x,y,\eta)$ under the assumption that the
local thermal equilibrium is reached (maintained) at (after) an initial
time $\tau_{0}$ =0.6 fm/$c$. Here $\tau$ and $\eta$ are the proper
time and the space-time rapidity, respectively. $x$ and $y$ are
transverse coordinates. In the transverse plane, the centers of two
colliding nuclei are located at $(x,y)=(b/2,0)$ and $(-b/2,0)$ before
the collision at an impact parameter $b$. Ideal hydrodynamics is
characterized by the energy-momentum tensor, \begin{equation}
T^{\mu\nu}=(e+P)u^{\mu}u^{\nu}-Pg^{\mu\nu},\end{equation}
 where $e$, $P$, and $u^{\mu}$ are energy density, pressure, and
local four velocity, respectively. We neglect the finite net-baryon
density which is small near the mid-rapidity at RHIC. For the high
temperature ($T>T_{c}=170$ MeV) QGP phase we use the equation of
state (EOS) of massless non-interacting parton gas ($u$, $d$, $s$
quarks and gluons) with a bag pressure $B$: \begin{eqnarray}
p=\frac{1}{3}(e-4B).\end{eqnarray}
The bag constant is tuned to be $B^{\frac{1}{4}}=247.19$\,MeV to
match pressure of the QGP phase to that of a hadron resonance gas
at critical temperature $T_{c}=170$\,MeV. A hadron resonance gas
model at $T<T_{c}$ includes all hadrons up to the mass of the $\Delta(1232)$
resonance. Our hadron resonance gas EOS implements chemical freeze-out
at $T_{\mathrm{ch}}=T_{c}=170$\,MeV, as observed in collisions at
RHIC \cite{BMRS01}. 

We assume that, at $\tau_{0}=0.6$ fm/$c$, the initial entropy distributions
is proportional to a linear combination of the number density of participants
(85\%) and binary collisions (15\%) \cite{Hirano06}. Centrality
dependence of charged particle multiplicity observed by PHOBOS \cite{PHOBOS_Nch}
has been well reproduced by full 3D hydrodynamics simulations with
the above setups~\cite{Hirano06}. In the following calculations,
hydrodynamic outputs at representative impact parameters $b=$ 3.2,
5.5, 7.2, 8.5, 9.7, and 10.8 fm are chosen for 0-10\%, 10-20\%, $\cdots$,
50-60\% centrality, respectively.

For convenience of the following calculations, we introduce $f_{{\rm \textrm{QGP}}}(\tau,x,y,\eta)$
as the fraction of the QGP phase in a fluid element. It is obvious
that $f_{{\rm \textrm{QGP}}}=1$ (0) in the QGP (hadronic) phase.
In the mixed phase, the fraction of the QGP is calculated via \begin{eqnarray*}
f_{\textrm{QGP}}\, e_{\textrm{QGP}}+(1-f_{\textrm{QGP}})e_{\textrm{had}}=e(\tau,x,y,\eta)\end{eqnarray*}
 with $e_{\textrm{QGP}}$ and $e_{\textrm{had}}$ being the energy
densities of the QGP phase and the hadron phase at $T=T_{c}$, respectively.

\section{Parton energy loss in a plasma}

\label{sec:3}
Hard parton production from primordial nucleus-nucleus
scattering is calculated as~\cite{Owens1987} \begin{eqnarray}
 &  & \frac{dN^{AB\rightarrow{\rm jet}}}{dyd^{2}\pt}=KT_{AB}(b)\sum_{abcd}\int dx_{a}dx_{b}G_{a/A}(x_{a},M^{2})\nonumber \\
 &  & \qquad\times G_{b/B}(x_{b},M^{2})\frac{\hat{s}}{\pi}\frac{d\sigma}{d\hat{t}}(ab\rightarrow cd)\delta(\hat{s}+\hat{t}+\hat{u})\label{eq:ABtojet}\end{eqnarray}
 where $T_{AB}(b)$ is the nuclear overlapping function at an impact
parameter $b$ for each centrality, $G_{a/A}(x_{a},M^{2})$ and $G_{b/B}(x_{b},M^{2})$
are parton distribution functions in nuclei $A$ and $B$. We take
MRST 2001 LO parton distributions in proton~\cite{MRST0201}. $K=2$
is chosen to take into account higher order contributions. These parameters
are chosen as to reproduce high $\pt$ pion data in $pp$ collisions
at RHIC, which will be discussed later. Nuclear shadowing effect and
EMC effect are taken into account through EKS98 scale dependent nuclear
ratios $R_{a}^{{\rm \textrm{EKS}}}(x,A)$~\cite{EKS98}. Isospin
of a nucleus with mass $A$, neutron number $N$, and proton number
$Z$ is corrected as follows: \begin{equation}
G_{a/A}(x)=\left[\frac{N}{A}G_{a/N}(x)+\frac{Z}{A}G_{a/P}(x)\right]R_{a}^{{\rm \textrm{EKS}}}(x,A).\label{eq:PdisA}\end{equation}
 The isospin mixture and nuclear shadowing eventually cause a decrease
of nuclear modification at high $\pt$ region.

We assume that all jets are produced at $\tau=1/Q\approx0$ with the
phase space distribution \begin{equation}
f_{0}(\vec{p},\vec{r})\propto\frac{dN}{d^{3}p}T_{A}\left(x-\frac{b}{2},y\right)T_{B}\left(x+\frac{b}{2},y\right)\delta(z)\label{eq:space dis}\end{equation}
 where $\vec{r}=(x,y,z)$ is the coordinate of a jet, $b$ is the
impact parameter, and $T_{A}$ and $T_{B}$ are thickness functions
of nuclei $A$ and $B$. 

When hard partons propagate in a hot dense matter, they may lose energy in the 
medium via radiation or elastic collisions with thermal partons\cite{bjorken82,Gy90,Baier97,BDMPS}. This is a quite hot issue and several different schemes have been employed for jet energy loss in QGP phase while the study of hard parton energy loss in hadronic phase is not so popular.
In this paper the BDMPS schemes is employed which considers the energy loss of hard parton in QGP phase via radiation. A free parameter $D$ is introduced by us to take into account of energy loss via other modes.

The total energy loss along the trajectory of a hard parton is calculated
as \begin{equation}
\Delta E(i,\vec{p}_{0},\vec{r}_{0})=D\int_{\tau_{0}}^{\infty}d\tau\epsilon(i,\tau,\bm{x}(\tau))\,\theta\big(f_{{\rm QGP}}(\tau,\bm{x}(\tau))\big),\label{eq:BDMPS}\end{equation}
where $\bm{x}(\tau)$ is the trajectory of parton, $f_{{\rm QGP}}(\tau,\bm{x}(\tau))$
is the fraction of the QGP phase at a position $(\tau,$$\bm{x}(\tau))$,
and $\theta$ is a step function, which gives $\theta(f_{{\rm QGP}})$
equal unity in the QGP and the mixed phases and zero in the hadron
phase. In the mixed phase, the thermal parton density will be diluted
with the factor $f_{{\rm QGP}}$, the fraction of the QGP phase. Here
$D$ is an adjustable parameter, $\epsilon(i,\tau,\bm{x}(\tau))$
is the energy loss per unit distance for a parton $i$ at a position
$(\tau,$$\bm{x}(\tau))$, given as~\cite{Baier97} \[
\epsilon(i,\tau,\bm{x}(\tau))=\alpha_{s}\sqrt{\mu^{2}E^{*}/\lambda_{i}}.\]
Note that the above quantities, $\ie$, temperature $T$, fluid velocity
$u_{\mu}$, parton densities $\rho_{i}$ and, in turn, mean free path
$\lambda_{i}$, depend on the location of the parton $\bm{x}(\tau)$
and can be obtained from full 3D hydrodynamics simulations discussed
in the previous section.

We first discuss pion production in proton-proton collisions. We calculate
neutral $\pi$-meson production assuming pQCD factorization, Eq.~(\ref{eq:ABtojet}),\begin{equation}
\frac{dN_{pp}^{\mathrm{\pi^{0}}}}{dyd^{2}\pt}=\sum_{c=g,q_{i}}\int dz_{c}\frac{dN^{pp\rightarrow c}}{dyd^{2}\pt^{c}}\frac{1}{z_{c}^{2}}D_{\pi^{0}/c}^{0}(z_{c},Q^{2}),\label{eq:jet-pi0}\end{equation}
 where $D_{\pi^{0}/c}^{0}(z_{c},Q^{2})$ is $\pi^{0}$ fragmentation
functions parameterized by Kniehl \textit{et al.}~\cite{KKPfrag}.
In the high $\pt$ region where the pQCD is expected to work, we reasonably
reproduce the experimental data with the above setup with $K=2$ and
$M=Q=\pt$. We use the $\pt$ spectrum as a reference spectrum in
the following calculations.

The effect of parton energy loss is taken into account through the
medium modified fragmentation function~\cite{XNWANG04} $D_{\pi^{0}/c}(z_{c},Q^{2},\Delta E_{c})$.
With a common value of the parameter $D=1.5$, we can reasonably reproduce
the PHENIX data~\cite{AuAu-pi-phenix} on $\pi^{0}$ jet quenching
in the high $\pt$ region at all centralities simultaneously. In the
following photon calculations, we always use the BDMPS energy loss
formula (\ref{eq:BDMPS}) with $D=1.5$.

\section{The different sources of direct photon production}

\label{sec:4}Possible sources of direct photons are considered in
the following.

\textit{Leading order contribution in Primordial NN scattering.} The
direct photon production via Compton scattering and quark-antiquark
annihilation can be calculated in perturbation theory using the conventional
parton distribution functions and the factorization hypothesis. Higher
order contributions are considered as a part of the so-called jet
fragmentation contribution and jet photon conversion which will be
discussed separately. Similar to Eq.~(\ref{eq:ABtojet}), the leading
order contribution to direct photon production in nucleus-nucleus
collisions reads \begin{eqnarray}
 &  & \frac{dN^{AB\rightarrow{\rm \gamma}}}{dyd^{2}\pt}=T_{AB}(b)\sum_{{\displaystyle ab}}\int dx_{a}dx_{b}G_{a/A}(x_{a},M^{2})\nonumber \\
 &  & \times G_{b/B}(x_{b},M^{2})\frac{\hat{s}}{\pi}\frac{d\sigma}{d\hat{t}}(ab\rightarrow\gamma+X)\delta(\hat{s}+\hat{t}+\hat{u})\label{eq:ABtogamma}\end{eqnarray}
where the elementary processes $ab\rightarrow\gamma+X$ are Compton
scattering $qg\rightarrow\gamma q$ and annihilation $q\bar{q}\rightarrow g\gamma$.

\emph{Thermal production.} In high energy nuclear collisions, the
density of secondary partons is so high that the quarks and gluons
rescatter and eventually thermalize to form a bubble of hot QGP. Thermal
photons can be produced via interaction between bulk particles during
the whole history of the evolution of hot matter. The interaction
of bulk particles and the bulk properties can be reflected by the
emission rate of direct photons in the QGP phase\cite{AMY2001} and
in the hadronic phase\cite{Rapp2004}, noted as $\Gamma=Ed^{3}R/d^{3}p$,
where $R$ is the number of photons emitted from a medium per unit
space-time volume with temperature $T$. Total yields of thermal photons
can be obtained by summing the emission rate over the space-time volume
as \begin{equation}
\frac{dN^{{\rm thermal}}}{dyd^{2}\pt}=\int d^{4}x\Gamma(p^{\mu}u_{\mu},T)\label{eq:E*a}\end{equation}
with $d^{4}x=\tau d\tau dxdyd\eta$, $p^{\mu}$ the photon's four
momentum in the laboratory frame and $u_{\mu}$ a local fluid velocity. 

\textit{Jet fragmentation.} At any stages of the evolution of a jet
(final state parton emission), there is a possibility of emitting
photons. The existence of a QGP affects the results because energetic
partons lose their energy prior to fragmentation. In this work, we
assume fragmentation of partons only outside the plasma, which is
similar to high $\pt$ hadron production from jet fragmentation.

In $pp$ collisions, there is no formation of QGP and no energy loss
of hard parton in QGP. Higher order contributions to direct photon
production can be considered as jet fragmentation and be calculated
as \begin{equation}
\frac{dN_{pp}^{\mathrm{frag}}}{dyd^{2}\pt}=\sum_{c=g,q_{i}}\int dz_{c}\frac{dN^{pp\rightarrow c}}{dyd^{2}\pt^{c}}\frac{1}{z_{c}^{2}}D_{\gamma/c}^{0}(z_{c},Q^{2}),\label{eq:jet-gamma0}\end{equation}
 with photon fragmentation functions $D_{\gamma/c}^{0}(z,Q^{2})$
being the probability for obtaining a photon from a parton $c$ which
carries a fraction $z$ of the parton's momentum. So $\pt^{c}=\pt/z_{c}$
is the transverse momentum carried by the parton $c$ before fragmentation
and $d^{3}p/E=z_{c}^{2}d^{3}p^{c}/E^{c}$. The effective fragmentation
functions for obtaining photons from partons can be calculated perturbatively.
We use the parameterized solutions by Owens~\cite{Owens1987}.

In case of heavy ion collisions where a QGP is formed, energy loss
of hard partons in a plasma should be taken into account. This can
be done via modified fragmentation functions~\cite{XNWANG04} \begin{eqnarray}
 &  & D_{\gamma/c}(z_{c},Q^{2},\Delta E_{c})\nonumber \\
 &  & =\left(1-e^{-\frac{L}{\lambda_{c}}}\right)\left[\frac{z_{c}'}{z_{c}}D_{\gamma/c}^{0}(z_{c}',Q^{2})+\frac{L}{\lambda_{c}}\frac{z_{g}'}{z_{c}}D_{\gamma/g}^{0}(z_{g}',Q^{2})\right]\nonumber \\
 &  & \quad+e^{-\frac{L}{\lambda_{c}}}\: D_{\gamma/c}^{0}(z_{c},Q^{2}),\end{eqnarray}
 with $z_{c}'=\pt/(\pt^{c}-\Delta E_{c})$ and $z_{g}'=(L/\lambda_{c})\,\pt/\Delta E_{c}$
being the rescaled momentum fractions carried by the parton $c$ and
the emitted gluons before fragmentation. $\lambda_{c}$ is mean free
path of the parton $c$ in the plasma, $L$ is the path length of
each parton traversing the plasma. Thus, in heavy ion collisions,
contributions from fragmentation become \begin{equation}
\frac{dN_{AB}^{{\rm frag}}}{dyd^{2}\pt}=\sum_{c=g,q_{i}}\int dz_{c}\frac{dN^{AB\rightarrow c}}{dyd^{2}\pt^{c}}\frac{1}{z_{c}^{2}}D_{\gamma/c}(z_{c},Q^{2},\Delta E_{c}).\label{eq:jet-gamma2}\end{equation}

\emph{Jet photon conversion with jet energy loss}. When hard partons
propagate in a plasma, they also collide with thermal partons and
produce direct photons via Compton process and the quark-antiquark
annihilation. We call this process jet-photon conversion, since it
is a conversion of a jet into a photon with almost the same momentum
as the one of originating jet parton. Contribution from the jet-photon
conversion is calculated by integration of conversion rate over the
space-time evolution of the hot and dense matter in the QGP phase
\begin{equation}
\frac{dN^{{\rm jpc}}}{dyd^{2}\pt}=\int\Gamma^{{\rm jpc}}(E^{*},T)f_{{\rm QGP}}(x,y,\eta,\tau)d^{4}x.\label{eq:jpc_int}\end{equation}
The photon production rate by annihilation and Compton scattering
of hard partons in the medium \cite{WangCY,Fries2005} \begin{equation}
\Gamma^{{\rm jpc}}(E^{*},T)=\frac{\alpha\alpha_{s}}{4\pi^{2}}\sum_{q}e_{q}^{2}f_{q}(\vec{p},x)T^{2}\left[\ln\frac{4E_{\gamma}^{*}T}{m_{{\rm th}}^{2}}-C\right]\label{eq:jpc_rate}\end{equation}
where $E^{*}$ is the photon energy in the local rest frame, $C=2.323$,
$m_{{\rm th}}^{2}=g^{2}T^{2}/6$, and the strong coupling $\alpha_{s}=g^{2}/4\pi$
being temperature dependent as in Eq.~(\ref{eq:alphasT}). $\alpha$
is the electromagnetic couplings, $e_{q}$ and $f_{q}(\vec{p},x)$
are the electric charge and the phase-space density of a hard parton
of flavor $q$. The phase space distribution of hard partons at $\tau$
is obtained by considering parton energy loss as \begin{eqnarray*}
f(\vec{p},x) & = & f(\vec{p},\vec{r},\tau)\\
 & = & \int d^{3}p_{0}f_{0}(\vec{p}_{0},\vec{r}-\vec{v}t)\delta(\vec{p_{0}}-\vec{p}-\vec{v}\Delta E)\end{eqnarray*}
 where $f_{0}(\vec{p},\vec{r})$ is the phase space distribution at
$\tau=0$ described in Eqs.~(\ref{eq:space dis}). The $\delta$-function
expression reflects the energy loss of a parton moving along a straight
line trajectory with $\vec{v}\equiv\vec{p}/E=\vec{p}_{0}/E_{0}$.
$\Delta E$ is the energy loss from $\tau_{0}$ to $\tau$ and calculated
similar to Eq.~(\ref{eq:BDMPS}) but replacing the upper limit of
integral $\infty$ with $\tau$.

\section{results and discussion}

\label{sec:5}

\begin{figure*}
\includegraphics[scale=0.8]{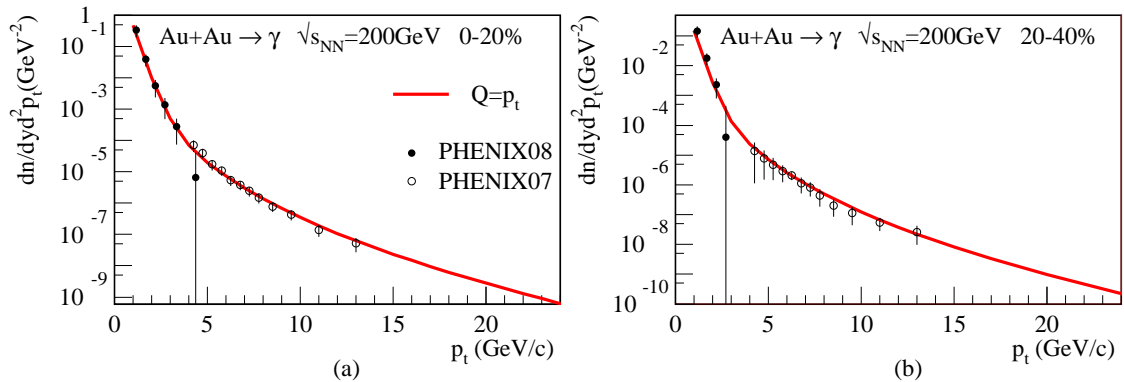}

\caption{\label{fig:c12-34} (Color Online) Direct photon production in Au+Au
collisions at centrality 0-20\% and 20-40\%. PHENIX data are shown
as open circles~\cite{PHENIX07PRL } and filled circles~\cite{PHENIX08g}.}
\end{figure*}

In Fig.~\ref{fig:c12-34}, the calculated $\pt$ spectra of direct
photons in Au+Au collisions at $\sqrt{s_{NN}}=200$ GeV at centrality
0-20\% and 20-40\% are compared to PHENIX data \cite{PHENIX07PRL ,PHENIX08g}.
The PHENIX data are reproduced within our multi-component model remarkably
well.

The nuclear modification factor $R_{AA}$ is obtained by dividing
a $\pt$ spectrum in nucleus-nucleus collisions by the $N_{\mathrm{coll}}$-scaled
$\pt$ spectrum in $pp$ collisions. The calculation of direct photons
in $pp$ collisions includes the leading order contribution plus fragmentation
contribution. In high $\pt$ regions, our result agrees with the PHENIX
data\cite{PHENIX07PRL },\cite{PHENIX08g} reasonably
well: So we use it to calculate nuclear modification. It also provides
a baseline calculation with the LO contribution and fragmentation
contribution in Au+Au collisions. 

\begin{figure*}
\includegraphics[scale=0.8]{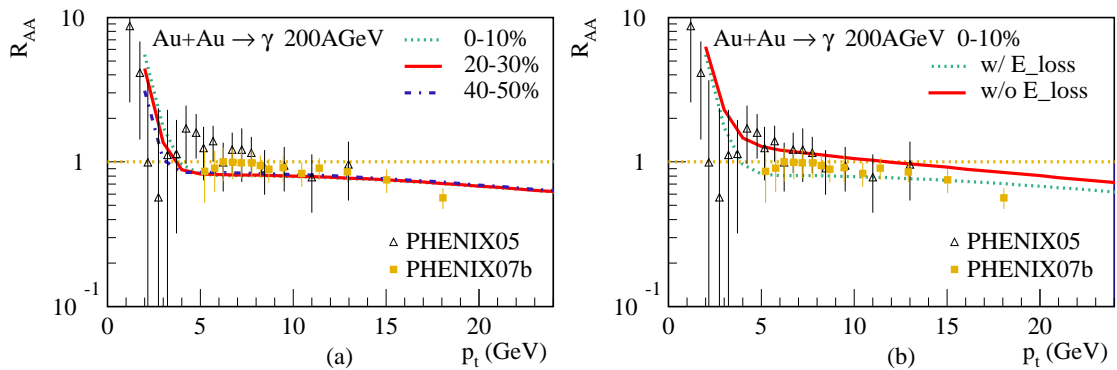}

\caption{\label{fig:Raa3} (Color Online) The nuclear modification factor
of direct photons in Au+Au collisions $R_{AA}$. Data for 0-10\% centrality
are from Ref.~\cite{PHENIX data} and Ref.~\cite{PHENIX07}. (a):
$R_{AA}$ at centrality 0-10\% (dotted line), 20-30\% (solid line),
and 40-50\% (dash-dotted line) respectively. (b): $R_{AA}$ at 0-10\%
centrality with energy loss (dotted line) and without energy loss
(solid line).}
\end{figure*}

Figure~\ref{fig:Raa3} shows how the nuclear modification factor
for direct photons, $R_{AA}$, depends on centrality and on energy
loss. Data for 0-10\% centrality are taken from Refs.~\cite{PHENIX data}
and \cite{PHENIX07}. Figure~\ref{fig:Raa3}(a) shows centrality
dependence of $R_{AA}$ compared to the PHENIX data. The three curves
are respectively 0-10\% (dotted line), 20-30\% (solid line), and 40-50\%
(dash-dotted line). 

$R_{AA}$ has a weak centrality dependence at high $\pt$ region.
This result is consistent with the observed phenomenon~\cite{PHENIX data}
that the $\pt$-integrated (for $\pt>6$~GeV/$c$) $R_{AA}$ of direct
photons is almost independent of collision centrality. Does this imply
a very weak effect from jet quenching? Figure~\ref{fig:Raa3}(b)
answers this question (here for the most central collisions): Comparing
calculations with (dotted line) and without energy loss (solid line),
one finds a difference of up to 40\%. So the effect of parton energy
loss is quite visible in the $\pt$ range between $4$ GeV/$c$ and
more than $20$ GeV/$c$. If we would do the $R_{AA}$ calculations
without energy loss, the difference between central and semiperipheral
collisions would be about 20\%, wheras the complete calculation gives
the same result for all centralities, within 5\%. We have to admit
that we talk about small effects, requiring experimental data with
relative errors of less than 5
to observe the effects. %
\begin{figure*}
\includegraphics[scale=0.7]{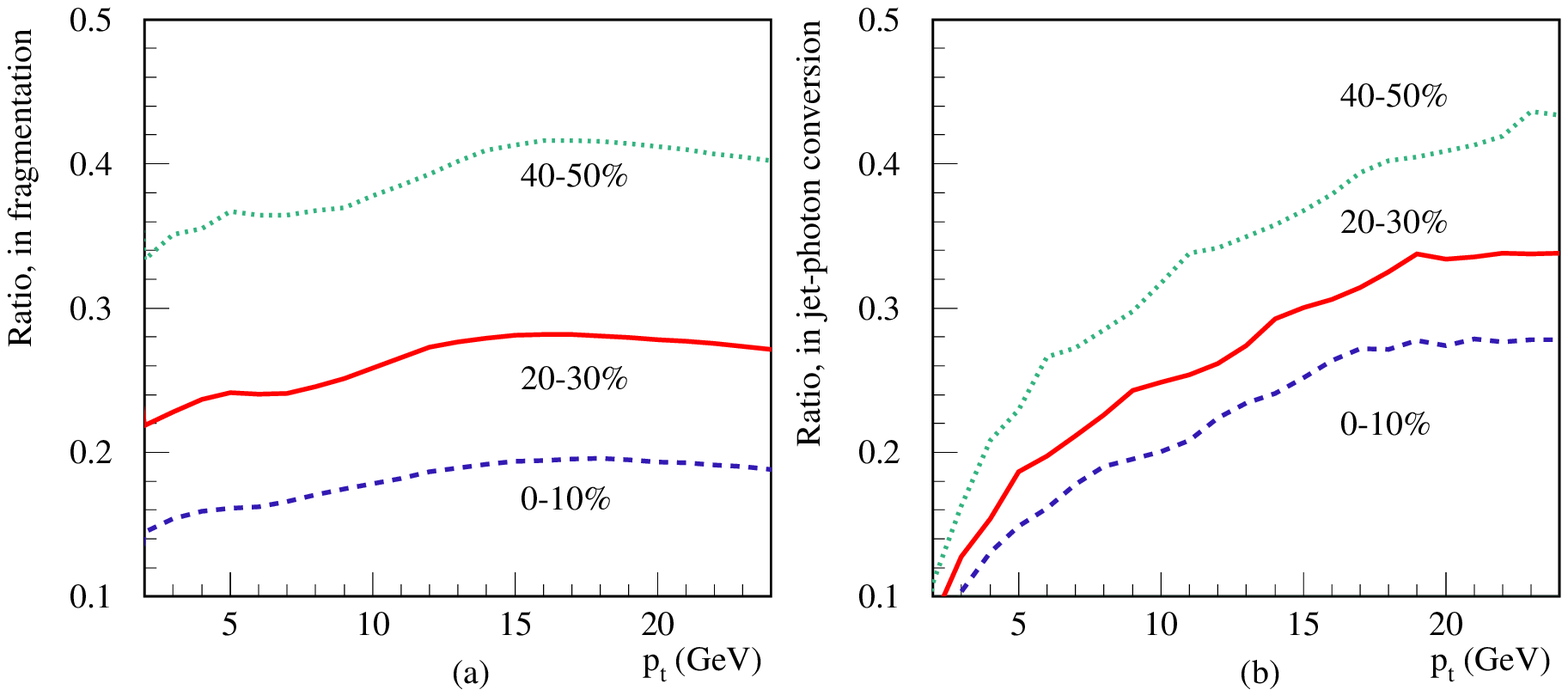}

\caption{\label{fig:Eloss} (Color Online) The ratio of the contribution with
energy loss to the one without, in fragmentation (a) and jet-photon
conversion (b).}
\end{figure*}

Parton energy loss in the plasma suppresses the fragmentation contributions
and jet-photon conversion. So we study the ratios of the contribution
with energy loss to the one without energy loss, as shown in Fig.~\ref{fig:Eloss}
((a) for fragmentation and (b) for jet-photon conversion). Energy
loss in the plasma depends on the path length of the hard parton inside
the plasma, which turns out to depend on the collision centrality.
We do see a similar centrality dependence of the suppression for $\pi^{0}$
(jet quenching effect) in fragmentation contributions and jet-photon
conversion.

\section{Conclusion}

\label{sec:6}

We calculated the centrality-dependence of $\pt$ spectra for direct
photons in Au+Au collisions at the RHIC energy, based on a realistic
data-constrained (3+1)-D hydrodynamic description of the expanding
hot and dense matter, a reasonable treatment of propagation of partons
and their energy loss, and a systematic consideration of main sources
of direct photons. In this study, four main sources are considered,
namely, leading order (LO) contribution from primordial elementary
scatterings, thermal radiation from the fluids, fragmentation from
hard partons, and jet photon conversion (JPC). Similar work~\cite{Turbide:2007mi}
has been done before the appearance of the most recent data~\cite{PHENIX08g}.
Our results agree nicely with the recent low $\pt$ data.

The role of jet quenching in the high $\pt$ region of direct photons
production has been checked via fragmentation photons and jet photon
conversion sources. For these two sources, the suppression of the
photon rate due to parton energy loss is significant in central Au+Au
collisions, and becomes less important towards peripheral collisions,
similar to the suppression for meson production. Since experimentally
one may separate isolated photons (LO+JPC) and associate photons (fragmentation
photons), our prediction may be tested in the future.

Considering the total yields of direct photons, the contribution from
fragmentation and conversion are small, contributing between 5\% and
10\%. However, parton energy loss plays nevertheless an important
role: Without it, these second order effects would contribute significantly.
Without jet quenching, the nuclear modification factors $R_{AA}$
would depend visibly on the centrality of the collisions. A strong
energy loss is actually necessary to get the centrality scaling of
$R_{AA}$ in our calculation -- a centrality scaling which has observed
by the PHENIX collaboration. In this sense, properties of the bulk
matter affect the photon yields at intermediate values of $\pt$,
via the parton energy loss.

\begin{acknowledgments}
This work is supported by the Natural Science Foundation of China
under the project No. 10505010 and MOE of China under project No.~IRT0624.
The work of T.H. was partly supported by Grant-in-Aid for Scientific
Research No.~19740130. 
\end{acknowledgments}

\end{document}